# Hot Carrier-Assisted Intrinsic Photoresponse in Graphene

Nathaniel M. Gabor[1], Justin C. W. Song[1,2], Qiong Ma[1], Nityan L. Nair[1], Thiti Taychatanapat[1,3], Kenji Watanabe[4], Takashi Taniguchi[4], Leonid S. Levitov[1], Pablo Jarillo-Herrero[1]

[1] *Department of Physics, Massachusetts Institute of Technology, Cambridge, MA 02139, USA.*

[2] *Harvard School of Engineering and Applied Sciences, Harvard University, Cambridge, MA 02138, USA.*

[3] *Department of Physics, Harvard University, Cambridge, MA 02138, USA.*

[4] *National Institute for Materials Science, Namiki 1-1, Tsukuba, Ibaraki 305-0044, Japan.*

**Abstract**

Graphene is a new material showing high promise in optoelectronics, photonics, and energy-harvesting applications. However, the underlying physical mechanism of optoelectronic response has not been established. Here, we report on the intrinsic optoelectronic response of high-quality dual-gated monolayer and bilayer graphene p-n junction devices. Local laser excitation at the p-n interface leads to striking six-fold photovoltage patterns as a function of bottom- and top-gate voltages. These patterns, together with the measured spatial and density dependence of the photoresponse, provide strong evidence that non-local hot-carrier transport, rather than the photovoltaic effect, dominates the intrinsic photoresponse in graphene. This novel regime, which features a long-lived and spatially distributed hot carrier population, may open the doorway for optoelectronic technologies exploiting efficient energy transport at the nanoscale.

Recent studies of graphene's remarkable optical, thermal, and electronic properties have attracted immense interest for its use as a novel optoelectronic material (*1*). Charge carriers in graphene exhibit an electron-hole symmetric light-like energy dispersion, spanning a large range (~1 eV) of energies, and a zero band gap with vanishing density of states at low energies (*1, 2*). These electronic properties are directly manifested in diverse aspects of graphene's photoresponse (*3-9*), which has been the focus of extensive research for optoelectronic, photonic and energy-harvesting applications.

The photoresponse of semiconductors, which determines the performance of optoelectronic devices, is governed by the energy relaxation pathways of photoexcited electron-hole (e-h) pairs: energy transferred to the lattice is lost as heat, while energy transported through charge carriers may be used to drive an optoelectronic circuit (*10*). Nanoscale systems can offer various ways to control energy relaxation pathways, potentially resulting in more efficient device operation. Novel relaxation pathways have indeed been demonstrated in nanocrystal quantum dots and carbon nanotubes using electron confinement (*11, 12*). In graphene, energy relaxation pathways are strongly altered by the vanishing electronic density of states (*13-15*). After initial relaxation of photo-excited carriers due to electron-electron scattering and optical phonon emission, electron-lattice energy relaxation can be quenched (*13, 15*). This creates a bottleneck that limits further energy redistribution into the lattice. With electron-to-lattice energy relaxation quenched, a novel transport regime is reached in which thermal energy is redistributed solely among electronic charge carriers. The photo-generated carrier population remains hot while the lattice stays cool.

In graphene, hot carriers should play a key role in optoelectronic response (*15*), yet no measurements have clearly determined the photocurrent generation mechanism. Numerous initial photocurrent studies suggested that photocurrent generated at graphene-metal contacts (*3-6*) results from the photovoltaic (PV) effect, in which a built-in electric field accelerates photogenerated charge carriers to the electronic contacts. More recent studies have looked at monolayer-bilayer interfaces (*8*) and highly controlled chemical potential gradients at gate voltage-controlled p-n interfaces (*9*), and indicate that photothermoelectric effects may play an important role.

In this work, we report optoelectronic transport measurements of gate voltage-controlled graphene p-n junction devices in the presence of local laser excitation that determine the intrinsic photoresponse. Our measurements unambiguously indicate that hot electronic carriers dominate graphene's intrinsic optoelectronic response, at temperatures ranging from room temperature down to 10 K and in the linear optical power regime. The hot carrier regime manifests as a strong photothermoelectric effect that results in a striking six-fold photovoltage pattern as a function of gate voltages. Additionally, the spatial and charge density dependence of the optoelectronic response establishes a strong connection between thermal energy transport and electronic charge carriers.

We study p-n junctions both in monolayer and bilayer graphene devices, integrated with global bottom gates and local top gates as shown in Figure 1A and described in (*16*). Tuning the bottom and top gate voltages, $V_{BG}$ and $V_{TG}$ respectively, allows independent control of the Fermi energy $E_F$, and thus carrier density of electrons (n-type carriers) and holes (p-type carriers) in each region (*17-19*). By applying voltages of opposite polarity on $V_{BG}$ and $V_{TG}$, we can form a p-n junction at the interface of p- and n-type regions in a single graphene sample (Figure 1C bottom).

We first characterize our devices via electronic transport measurements of the resistance $R$ as a function of $V_{BG}$ and $V_{TG}$. Figure 1B shows $R$ vs. $V_{BG}$ and $V_{TG}$ for a monolayer graphene (MLG) p-n junction. The resistance exhibits four characteristic regions, a distinctive behavior that reflects gate voltage-tunable charge density and a sharp density gradient at the p-n interface (*17-19*). When the charge density $n_c$ is tuned near the neutrality point in each gated region, $R$ increases due to the low density of conduction carriers. When two gates are present, this results in two intersecting lines of relatively high resistance, and a maximum resistance $R_{MAX}$ at their intersection, the global charge neutrality point (CNP). The two lines divide the resistance map into four regions: p-n, n-n, n-p, and p-p, labeled according to the carrier doping induced in the bottom-gated and dual-gated regions (regions 1 and 2, respectively). Importantly, as the carrier density increases away from the CNP, the resistance decreases monotonically from $R_{MAX}$.

The dual-gated graphene p-n device is placed in a low temperature optical cryostat that combines electronic transport with mirror-controlled scanning laser excitation (Figure 1C) (*16*). Figure 1D shows a photocurrent image obtained by scanning a 1 micron-diameter focused laser spot ($\lambda$ = 850 nm) over the device at $V_{SD}$= 0 V while measuring the current $I$. At the sharply defined p-n interface, we observe a strong photocurrent that is an order of magnitude larger than the photocurrent at the contacts and increases linearly with optical power. The maximum photocurrent responsivity reaches 5 mA/W at low temperatures, larger than previously reported values at room temperature (*6, 9*). As a function of distance away from the sharp p-n interface, the photocurrent exhibits a triangular shape that extends several microns to the contacts (solid line in Figure 1E). As we will show, a triangular photocurrent profile indicates that charge carriers remain hot in order to reach the p-n interface and contribute to photocurrent.

Unlike the resistance in which we observe four gate voltage regions, optoelectronic transport measurements with the laser fixed at the p-n interface exhibit a striking photovoltage pattern with *six* regions of alternating photovoltage sign as a function of $V_{BG}$ and $V_{TG}$ (Figure 2). Since the photoconductance in graphene is negligible (*5*), the photovoltage established across the device can be determined from $V_{PH} = I_{PH}R$. If the photocurrent originates from the PV effect, in which the photovoltage is proportional to the electric field induced at the p-n interface, then we would expect only a single photovoltage sign change and a monotonic dependence on increasing charge carrier density (*15*). In remarkable contrast, the photovoltage $V_{PH}$ as a function of $V_{BG}$ and $V_{TG}$ (Figure 2A) clearly exhibits multiple sign changes. Moreover, the photovoltage line traces extracted along the high resistance ridges (gray lines), along which charge density increases away from the CNP, exhibit strong non-monotonic gate voltage-dependence (Figure 2A left and bottom). Below we discuss detailed optoelectronic and thermoelectric measurements that probe the origins of this behavior.

The shape of the photovoltage node lines ($V_{PH} = 0$ lines) indicates that while $R$ decreases monotonically away from $R_{MAX}$ at the CNP, the photovoltage shows non-monotonic dependence on gate voltages. In the PV effect, due to the changing direction of the electric field as the relative carrier densities are changed in regions 1 and 2, photovoltage would change sign across a single node line. This line, along which the carrier density in the two regions is equal (white dashed line in Figure 2A), divides the p-p and n-n regions (*15*). Figure 2B shows the photovoltage nodes determined by plotting the absolute value of photovoltage $|V_{PH}|$ vs. $V_{BG}$ and $V_{TG}$. We observe three photovoltage nodes that deviate significantly from lines of high resistance. Near the CNP, where $R$ decreases monotonically away from $R_{MAX}$, the photovoltage displays a vertically oriented z-shaped nodal line and two points towards which the three nodal lines converge (arrows in Figure 2B).

Such non-monotonic behavior of transport properties is reminiscent of the sign change and non-monotonic charge density dependence of the thermoelectric voltage in graphene (*20-23*). Importantly, thermoelectric transport measurements on the same sample show that the thermovoltage $V_T$ measured as a function of $V_{BG}$ exhibits similar features to the photovoltage nodes along the high resistance ridges. By laser-heating the gold contact far from the graphene-metal junction (marked by a triangle in Figure 1D), we introduce a fixed un-calibrated temperature gradient across the device and measure the resulting thermovoltage (see also (*16*)). Figure 2C shows $V_T$ and $R$ as a function of $V_{BG}$. While $R$ decreases monotonically away from the CNP, $V_T$ exhibits clear non-monotonic gate voltage-dependence with peaks that correspond closely to the gate voltages at which the $V_{PH}= 0$ lines converge near the CNP (arrows Figure 2B).

We attribute the intricate photovoltage pattern in graphene to the photothermoelectric (PTE) effect. After optical excitation, e-h pairs decay rapidly (~10-150 fs) into a thermal distribution of electronic carriers (*24, 25*) with a local effective temperature $T_e$ that remains out of equilibrium with the lattice. The difference in temperature with its surroundings results in a thermal current that is accompanied by a charge current, the magnitude and sign of which is sensitive to carrier type and density (i.e. a p-type sample exhibits charge current with an opposite sign to that of an n-type sample). In un-biased homogeneous graphene, an isotropic thermal current density flowing away from the excitation spot is accompanied by a similar isotropic charge current density resulting in zero net current.

However, in a p-n junction this symmetry is broken, and photo-excitation can result in non-zero net PTE current. The sign and magnitude of PTE voltage resulting from this current depends on the Seebeck coefficient (*8, 26*) in each region and can be written as

$$V_{PTE} = (S_2 - S_1)\Delta T \tag{1}$$

where *S* is thermoelectric power (Seebeck coefficient) in the bottom- and dual-gated regions, and *ΔT* taken at the p-n interface is proportional to the electron temperature difference between the area of optical excitation and its surroundings (*15*). From the Mott formula (*26*), we can write *S* as

$$S = \frac{\pi^2 k_B^2 T}{3e} \frac{1}{R} \frac{dR}{dV_G} \frac{dV_G}{dE}\bigg|_{E=E_F} \tag{2}$$

where *T* is the sample temperature, $k_B$ is the Boltzmann constant, and $E_F$ is the Fermi energy (*8, 20-23*). While *R* decreases monotonically with $V_G$ away from the CNP, *S* changes non-monotonically (see schematic Figure 2D). The non-monotonic dependence of $S_1$ and $S_2$ results in multiple sign reversals for the quantity ($S_1$ - $S_2$), which occur along 3 nodal lines. This gives rise to the six-fold pattern (Figure 2A) (*15*). The six-fold pattern thus serves as a litmus test for the photothermoelectric effect.

We can gain more direct insight into the origin of the six-fold photovoltage patterns by extracting the two components of the PTE voltage individually, using the Fourier analysis technique (*16*). This is possible because $S_1$ and $S_2$ depend on different combinations of $V_{BG}$ and $V_{TG}$. The Fourier transform of the data, displayed in Figure 3A, clearly exhibits a pair of streaks. The orientations of these streaks, horizontal (labeled $N_B$) and diagonal (labeled $N_T$), are described by the capacitance matrix of regions 1 and 2 (see (*27*)). The relationship between the streaks and individual properties of the two regions can be clarified by masking one of the streaks and performing an inverse Fourier transform of the other streak. This gives two contributions, each behaving as a function of density in the corresponding regions, $V_{PH}[N_B]$ and $V_{PH}[N_T]$ (see Figure 3B top panels). Furthermore, the non-monotonic density dependence of each contribution is similar to the behavior of the Seebeck coefficient given by the Mott formula (see Figure 3B bottom panel). The sum $V_{PH}[N_B] + V_{PH}[N_T]$, reproduces the photovoltage pattern (Figure 3A lower inset). The non-monotonic behavior of each term produces polarity reversal on three nodal lines, giving rise to a six-fold pattern, as discussed above.

Six-fold patterns are also observed in the bilayer graphene (BLG) photoresponse. Composed of two Bernal-stacked sheets of monolayer graphene, BLG has a similar conical band structure to MLG at high energies, though with hyperbolic bands that touch at zero energy, resulting also in the absence of a band gap (*28-30*). Figure 4 shows the photoresponse of a BLG p-n junction photodetector. Using a top-gated configuration, two back-to-back p-n junctions can be tuned by gate voltages. The BLG resistance behaves similarly to the MLG p-n junction, showing four gate voltage regions separated by lines of high resistance (gray dashed lines in Figure 4C)(see also (*16*)). Strong photoresponse is observed in the p-n-p configuration, with photocurrent that peaks at the p-n junctions formed near the top gate edge (Figure 4B). The six-fold pattern in photovoltage vs. $V_{BG}$ and $V_{TG}$ indicates the presence of strong PTE effect (Figure 4C).

The hot carrier-assisted photoresponse in graphene is highly non-local, characterized by a long-lived and spatially distributed hot electronic population resulting from local excitation. The PTE photoresponse (Equation (1)) depends directly on the electronic temperature, which can be strongly enhanced due to long scattering times $\tau$ between hot electrons and low-energy phonons. The spatial profile of the electronic temperature can be described by a heat equation including a decay term that allows energy to leave the system (*15, 16*). Since the cooling length $\xi \sim \tau^{1/2}$ (*26*), long scattering times manifest as long cooling lengths. We fit the photocurrent profile (dashed line in Figure 1D) with the solutions of the heat equation and obtain excellent fits for $\xi > 2$ microns (*16*). While the cooling length may exceed 2 microns, the dimensions of the device allow us to establish only a lower bound, and not the exact cooling length. In contrast, if phonon-mediated substrate cooling dominated, heat energy would be dissipated efficiently downward into the substrate leading to lateral cooling lengths shorter than the beam spot size.

Graphene's intrinsic photoresponse also exhibits charge density dependence that confirms hot carrier-assisted transport of thermal energy. For excitation fixed at the p-n interface, the steady state value of $\Delta T$ at the laser spot is determined both by the energy flow into the spot due to laser power and heat flow out of the spot induced by the increased temperature (*16*). The heat flow out of the spot can be transported either through region 1 or region 2 towards the heat sinks at the contacts, giving (in the limit of long cooling length)

$$\Delta T = dQ/dt \, / \, (K_1 + K_2) \qquad (3)$$

where $dQ/dt$ is the rate of heat entering the system and $K_1$ ($K_2$) is the thermal conductance of region 1 (region 2). If heat flow is confined exclusively to the electronic degrees of freedom, then the thermal conductance obeys the Wiedemann-Franz relation $K = \pi^2 k_B^2 T / 3 e^2 \times (1/R)$, resulting in a charge density dependent $\Delta T \sim 1/K \sim R$ (*26*). In Figure 3C, we compare the photovoltage with the thermoelectric voltage obtained from a fixed temperature difference to isolate the density dependence of $\Delta T$. To eliminate any density dependence originating from $S$, we take the ratio of the two, $V_{PH}[N_B]/V_T$ (*16*). The ratio $V_{PH}[N_B]/V_T$ in Figure 3C clearly exhibits a density dependence that mimics the device resistance. From Figure 2A, we can estimate the electronic temperature using the Seebeck coefficient measured by Zuev et al. (*22*). The obtained value $\Delta T \sim 10$ K, which increases linearly with optical power, is a significant fraction of the lattice temperature at $T = 40$ K. The strong density dependence peaking at the CNP clearly demonstrates the electronic origin of $\Delta T$ in Equation (1) and is consistent with the density dependence expected from the Wiedemann-Franz relation.

Transport of hot electronic carriers results in a novel non-local transport regime that may make possible increased power conversion efficiency in energy-harvesting devices. At present, graphene is considered to be an excellent candidate for energy harvesting optoelectronics, in part due to the presence of high-responsivity photo-detection with high internal quantum efficiency (*5*, *15*). In this work, we have shown that hot carriers dominate the intrinsic photoresponse. Hot carrier-assisted photoresponse is predicted to improve the power conversion efficiency of energy harvesting devices significantly beyond standard limits (*31*). Our findings therefore promote the viability of graphene-based systems for energy harvesting applications, as long as new design principles are used in next-generation solar thermoelectric devices (*32*), which make use of hot carrier transport. This novel transport regime, which results from carriers that remain hot while the lattice stays cool, provides the ability to directly control the efficient transport of energy through its coupling to the electronic degree of freedom.

**References and notes**

1.  F. Bonaccorso, Z. Sun, T. Hassan, A. C. Ferrari. *Nature Photonics* **4**, 611-622 (2010).
2.  A. H. Castro Neto, F. Guinea, N. M. R. Peres, K. S. Novoselov, A. K. Geim. *Review of Modern Physics* **81**, 109-162 (2009).
3.  E. Lee, K. Balasubramanian, R. Weitz, M. Burghard, K. Kern. *Nature Nanotechnology* **3**, 486-490 (2008).
4.  J. Park, Y. H. Ahn, C. Ruiz-Vargas. *Nano Letters* **9**, 1742-1746 (2009).
5.  F. Xia, T. Mueller, Y. Lin, A. Valdez-Garcia, P. Avouris. *Nature Nanotechnology* **4**, 839-843 (2009).
6.  T. Mueller, F. Xia, P. Avouris. *Nature Photonics* **4**, 297-301 (2010).
7.  G. Nazin, Y. Zhang, L. Zhang, E. Sutter, P. Sutter. *Nature Physics* **6**, 870-874 (2010).
8.  X. Xu, N. M. Gabor, J. Alden, A. van der Zande, P. L. McEuen. *Nano Letters* **10**, 562-566 (2010).
9.  M. C. Lemme, et al. *ArXiv*:1012.4745v1 (2011).
10. S. M. Sze, *Physics of semiconductor devices*, 2$^{nd}$ ed. (Wiley, London, 1981).

11. V. Sukhovatkin, S. Hinds, L. Brzozowski, E. H. Sargent. *Science* **324**, 1542-1544 (2009).

12. N. M. Gabor, Z. Zhong, K. Bosnick, J. Park, P. L. McEuen. *Science* **325**, 1367-1371 (2009).

13. R. Bistritzer, A. H. MacDonald. *Physical Review Letters* **102**, 206410 (2009).

14. T. Winzer, A. Knorr, E. Malic. *Nano Letters* **10**, 4839 (2010).

15. J. C. Song, M. S. Rudner, C. M. Marcus, L. S. Levitov, *ArXiv*:1105.1142v1 (2011).

16. Information on Materials and Methods is available online.

17. B. Huard, et al. *Physical Review Letters* **98**, 236803 (2007).

18. J. Williams, L. DiCarlo, C. M. Marcus. *Science* **317**, 638-641 (2007).

19. B. Ozyilmaz, et al. *Physical Review Letters* **99**, 166804 (2007).

20. E. H. Hwang, E. Rossi, S. Das Sarma. *Physical Review B* **80**, 235415 (2009).

21. P. Wei, W. Bao, Y. Pu, C. N. Lau, J. Shi. *Physical Review Letters* **102**, 166808 (2009).

22. Y. Zuev, W. Chang, P. Kim. *Physical Review Letters* **102**, 096807 (2009).

23. J. Checkelsky, N. P. Ong. *Physical Review B* **80**, 081412(R) (2009).

24. P. George, et al. *Nano Letters* **8**, 4248-4251 (2008).

25. D. Sun, et al. *Physical Review Letters* **101**, 157402 (2008).

26. N. W. Ashcroft, N. D. Mermin, *Solid State Physics* (Thomson Learning Inc, USA, 1976).

27. In the top gated region, both $V_{BG}$ and $V_{TG}$ contribute to $n_c$, resulting in the slope of the diagonal line $m=\Delta V_{TG}/\Delta V_{BG}= C_{BG}/C_{TG} \sim 0.05$ where $C_{BG}$ ($C_{TG}$) is the bottom (top) gate capacitance to graphene. $N_B$ and $N_T$ in the Fourier transform agree with the capacitance matrix, i.e. they are perpendicular to the minimum resistance lines (grey dashed lines) of Figure 2A.

28. E. McCann. *Physical Review B* **74**, 161403 (2006).

29. T. Ohta, A. Bostwick, T. Seyller, K. Horn, E. Rotenberg. *Science* **313**, 951-954 (2006).

30. E. Castro, et al. *Physical Review Letters* **99**, 216802 (2007).

31. R. T. Ross, A. J. Nozik. *Journal of Applied Physics* **53** (5), 3813-3818 (1982).

32. D. Kraemer, et al. *Nature Materials* **10**, 532-538 (2011).

**Acknowledgements** We thank M. Baldo, P. McEuen, P. Kim, and A. Yacoby for valuable discussions. This work was supported by the Air Force Office of Scientific Research, NSF Early Career Award and by the Packard Foundation. Sample fabrication was performed at the NSF funded MIT Center for Materials Science and Engineering and Harvard Center for Nanoscale Science.

**Figure legends**

**Figure 1 Device geometry, band structure, and optoelectronic characteristics of the graphene p-n junction.** (**A**) Optical microscope image of the dual gated device incorporating boron nitride top gate dielectric. White lines mark the boundaries of graphene, dark blue is boron nitride gate dielectric. (**B**) Resistance vs. $V_{BG}$ and $V_{TG}$ at $V_{SD}$=1.4 mV and $T$=175 K. Four regions are labeled according to carrier doping, p- type or n-type, in the bottom- and top-gated regions, respectively. (**C**) **Top** Experimental schematic of the graphene p-n junction in the scanning photocurrent microscope. **Bottom** Schematic of monolayer graphene's conical band structure in the p-n junction showing electron (blue) and hole bands (red). The dotted line represents the electron Fermi energy. (**D**) Spatially resolved photocurrent map at $T$=40 K with laser $\lambda = 850$ *nm* and optical power =50 μW ($V_{BG}$= -5 V, $V_{TG}$= 2 V, $V_{SD}$= 0 V). White lines mark location of gold contact and gate electrodes. Scale bar 5 microns. (**E**) Line trace (solid line) of the photocurrent map (taken at dashed line in (**D**)) as a function of distance from the p-n interface. The dashed line is a fit to the data (discussed in text).

**Figure 2 Gate-dependent photovoltage and thermoelectric response of the graphene p-n junction.** (**A**) p-n interface photovoltage vs. $V_{BG}$ and $V_{TG}$ at $V_{SD}$=0 V and $T$=40 K (measured at diamond in Figure 1D). Gray dashed lines are lines of high resistance from transport characteristics. Along the white dashed line, charge density is equal in region 1 and 2 (as described in text). **Left, Bottom**, line traces from the photovoltage map taken along the vertical and diagonal gray lines, respectively. (**B**) Absolute value of photovoltage near the CNP (labeled). (**C**) Resistance $R$ and thermovoltage $V_T$ vs. $V_{BG}$ at $V_{SD}$=0 V, $V_{TG}$=2.0 V and $T$=40 K (optical

power =1 mW, laser at triangle in Figure 1D). (**D**) Schematic of the Seebeck coefficient as a function of Fermi energy for typical resistance characteristics of graphene (**inset**).

**Figure 3 Fourier analysis of the photovoltage response in the graphene p-n junction**. (A) Fourier transform $|F\{V_{PH}\}|$ of photovoltage vs. $V_{BG}$ and $V_{TG}$ from Figure 2A. $N_B$ and $N_T$ label two components of the Fourier transform masked to extract the bottom- and top- gated photovoltage components. **Inset top**, photovoltage map from Figure 2A. **Inset bottom**, sum of the individual Fourier components of the photovoltage. (**B**) Bottom-gated and top-gated photovoltage components $V_{PH}[N_B]$ and $V_{PH}[N_T]$ as a function of $V_{BG}$ and $V_{TG}$ calculated by inverse Fourier transforming along the masked direction $N_B$ and $N_T$. **Bottom**, photovoltage line traces from $V_{PH}[N_B]$ and $V_{PH}[N_T]$. The $V_{BG}$ axis has been shifted by the CNP voltage for $V_{PH}[N_B]$. (**C**) Ratio of the photovoltage $V_{PH}[N_B]$ to the thermovoltage $V_T$, and resistance as a function of $V_{BG}$ measured along the white dashed line in Figure 2A (see also (*16*)).

**Figure 4 Spatial photocurrent and gate-dependent photovoltage in the bilayer graphene p-n-p photodetector.** (**A**) Optical microscope image of the BLG device. White lines mark the boundaries of BLG. (**B**) Photocurrent map at $T = 40$ K at $\lambda=850$ *nm* and optical power =200 µW ($V_{BG}=15$ V, $V_{TG}=10$ V, $V_{SD}=0$ V). White lines indicate location of gold contact electrodes. Scale bar 4 microns. (**C**) Photovoltage at the n-p junction (diamond in Figure 3C) vs. $V_{BG}$ and $V_{TG}$ at $T=40$ K and $V_{SD}=0$ V. Gray dashed lines are lines of high resistance taken from transport characteristics at $T=40$ K.

# Figures

Figure 1

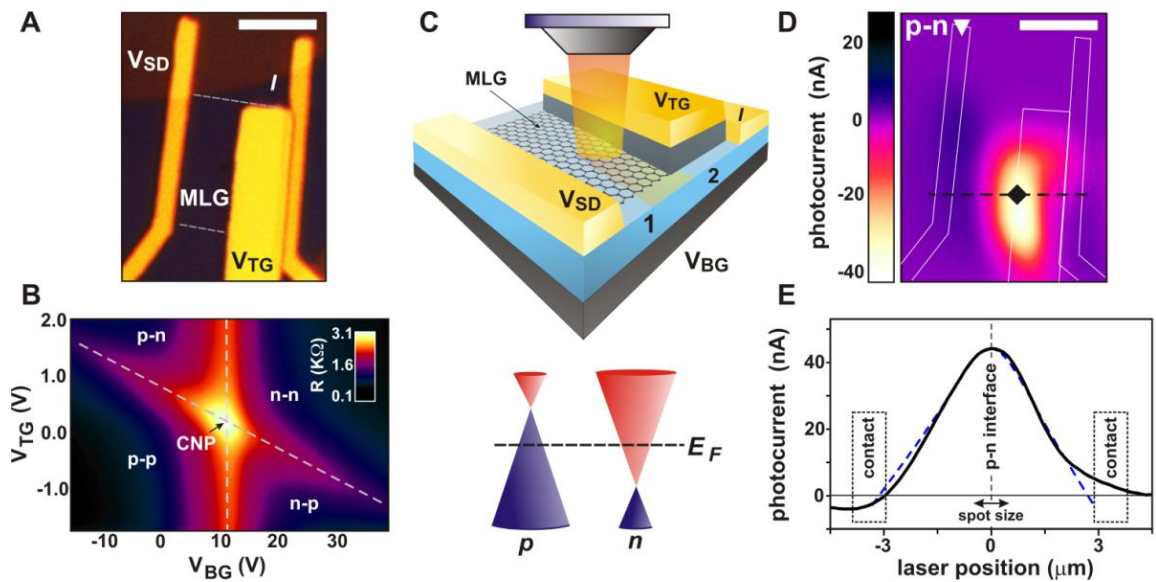

Figure 2

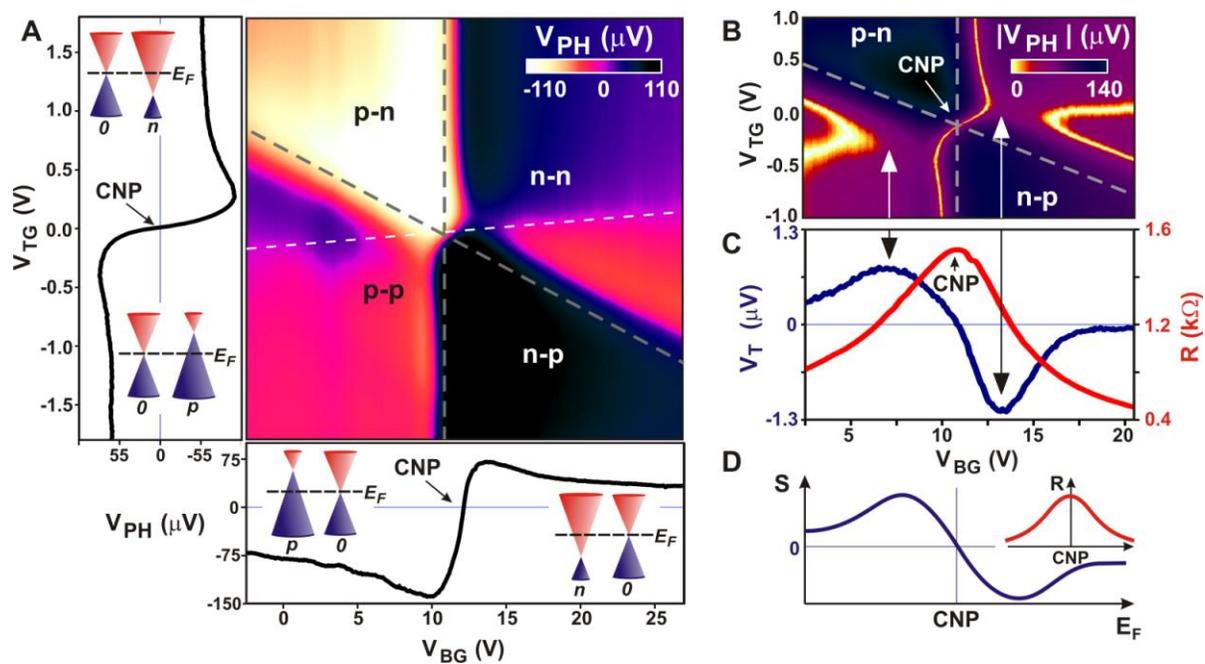

Figure 3

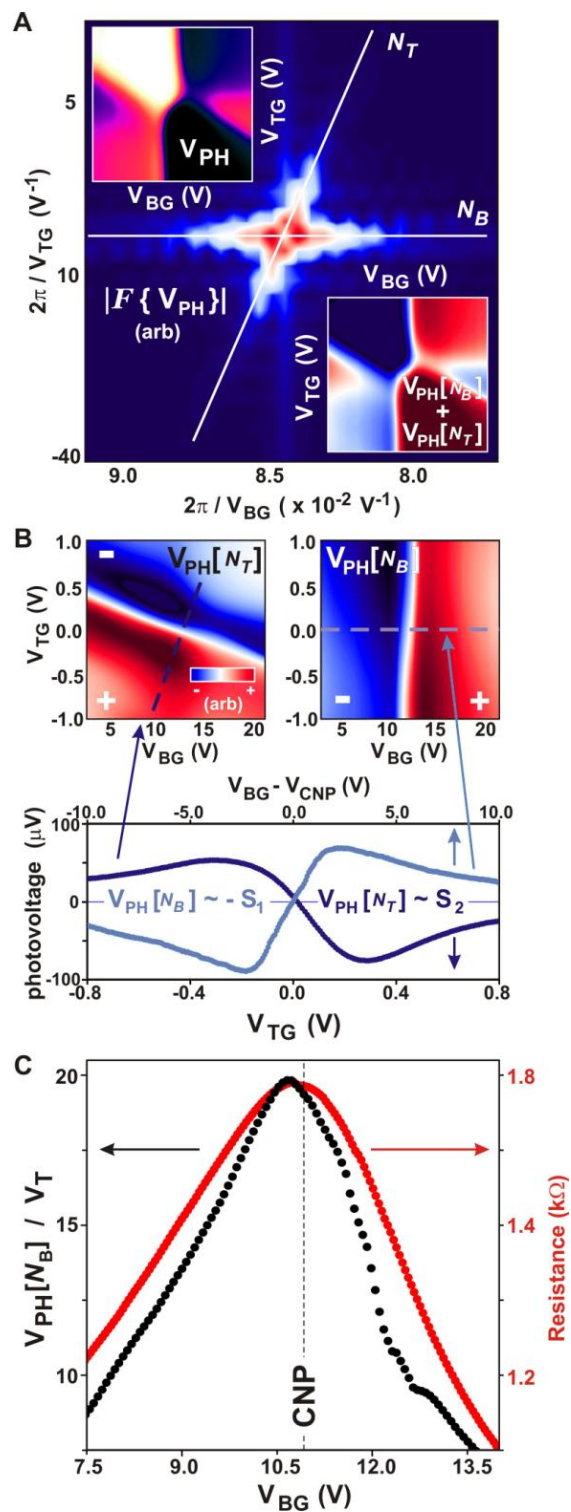

Figure 4

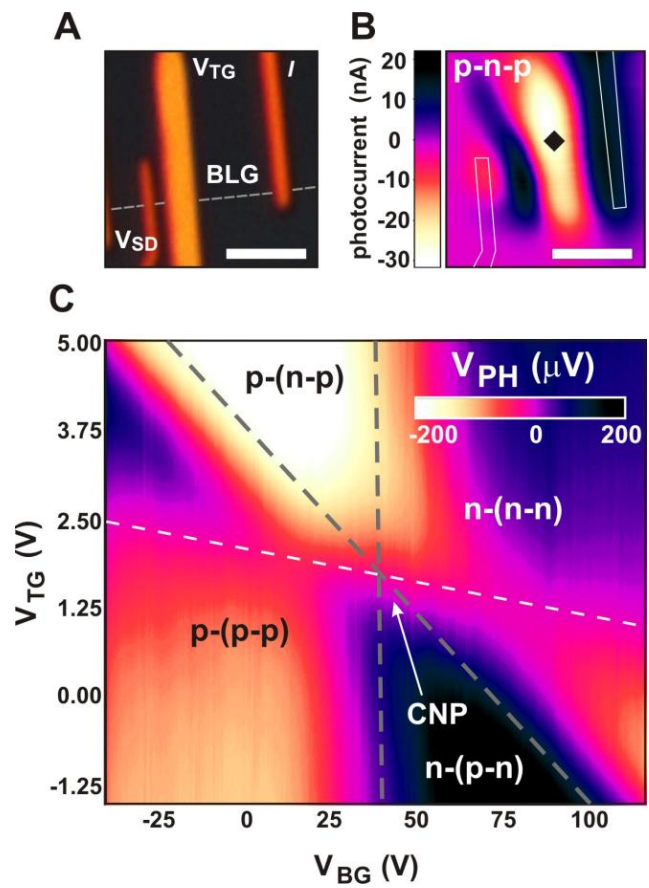